\documentclass[floatfix,aps,prl,twocolumn,amsmath,amssymb,groupedaddress]{revtex4}
\usepackage{graphicx}

\begin{document}

\title{Polarization conversion and "focusing" of light propagating through a small chiral hole in a metallic screen}

\author{A.\,V.~Krasavin}
\email[]{avk@soton.ac.uk}
\author{A.\,S.~Schwanecke}
\author{N.\,I.~Zheludev} 
\affiliation{EPSRC NanoPhotonics Portfolio Centre, School of
Physics and Astronomy, University of Southampton, SO17 1BJ, UK}
\author{M.~Reichelt}
\author{T.~Stroucken}
\author{S.\,W.~Koch}
\affiliation{%
Department of Physics and Material Sciences Center,
Philipps-University, Renthof 5, D-35032 Marburg, Germany}
\author{E.\,M.~Wright}
\affiliation{Optical Sciences Center and Department of Physics,
University of Arizona, Tucson, AZ 85721, USA}

\date{\today}

\begin{abstract}
Propagation of light through a thin flat metallic screen
containing a hole of twisted shape is sensitive to whether the
incident wave is left or right circularly polarized. The
transmitted light accrues a component with handedness opposite to
the incident wave. The efficiency of polarization conversion
depends on the mutual direction of the hole's twist and the
incident light's wave polarization handedness and peaks at a
wavelength close to the hole overall size. We also observed a
strong transmitted field concentration at the center of the chiral
hole when the handedness of the chiral hole and the wave's
polarization state are the same.
\end{abstract}

\maketitle

Metallic films structured on the nanoscale show a range of unusual
properties such as extraordinarily high transmission of light
through metallic films perforated with round, rectangular and
C-shaped holes \cite{Ebbesen,Abajo,Matteo} and strong linear
birefringence for asymmetrical openings \cite{Elliott,Degiron}.
Metallic films with arrays of chiral (twisted) holes, i.e. holes
that are not their mirror images over any line in the plane of the
film, are of special interest. They show polarization effects in
diffraction that are enantiomer-sensitive, i.e. depend on whether
the structural elements of the array are twisted clockwise or
anticlockwise \cite{Papacostas}, while their microscope images in
polarized light display unusual symmetries \cite{Schwanecke}. Here
we show that propagation of light through a metallic screen
containing a hole of twisted shape is sensitive to whether the
incident wave is left (LCP) or right circularly polarized (RCP). The
transmitted light accrues a component with handedness opposite to
the incident wave intensity which peaks at a wavelength close to the
hole overall size. We also observed a strong concentration of the
transmitted field at the center of the chiral hole when the
handedness of the chiral hole and the wave's polarization state are
the same.
\begin{figure*}
    \includegraphics[width=150mm]{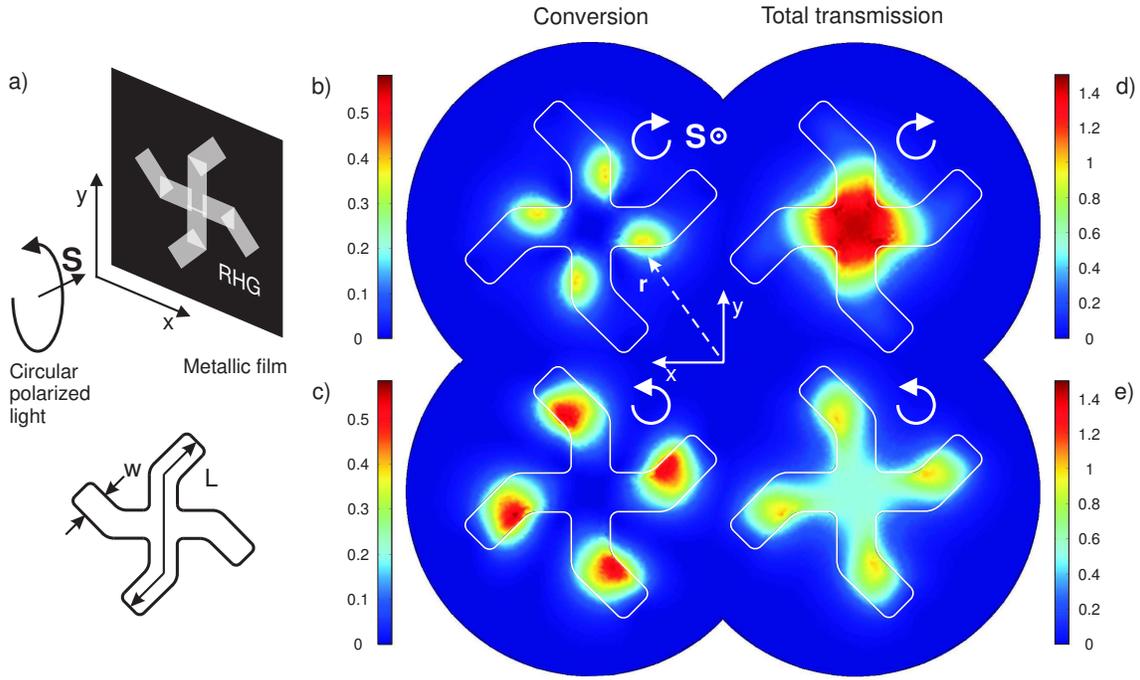}\\
  \caption{\label{Fig1}(a) General arrangement of the numerical simulations. \textit{Left field maps:} intensity
$(S_0^{tr}(\textbf{r})\pm S_3^{tr}(\textbf{r})) / (2
S_0^{in}(\textbf{r}))$ of the polarization converted component of
transmitted radiation for (b) RCP and (c) LCP incident waves.
\textit{Right field maps}: intensity $S_0^{tr}(\textbf{r}) /
S_0^{in}(\textbf{r})$ of total transmitted radiation for (d) RCP
and (e) LCP incident waves. For all field maps the incident wave
propagates in the direction $\bm{S}$ of the reader. The arrows
illustrate the rotational direction of the electric field vector
of the incident wave in the plane of the structure.}
\end{figure*}
\begin{figure}
    \includegraphics[width=75mm]{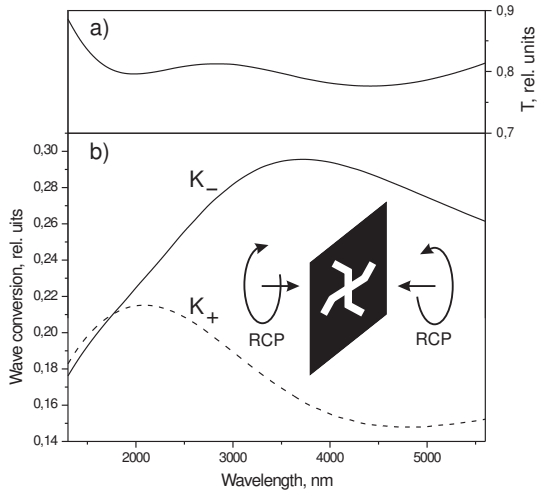}\\
  \caption{\label{Fig2}(a) Wavelength dependence of the total transmission $T$ through
  the gammadion opening; (b) wavelength dependence of the total polarization
  conversion efficiencies $K_{\pm}$ for
  incident waves with the same (dashed line, $K_{+}$) and opposite (solid line, $K_{-}$) handedness as the gammadion.
  Inserted figure shows that the perceived sense of twist of a hole in a flat screen
   depends on from which side of the screen it is observed. The polarization conversion
   efficiency for a wave of a given handedness depends
on whether it falls on the screen from one side or another.
  }
\end{figure}

For our study of chirality sensitive light propagation through a
chiral hole  we took a chiral opening of gammadion shape as an
example. The gammadion arms were bent at 45 degrees, as shown on
Fig.~\ref{Fig1}(a), to make a clock-wise twisted structure, with a
positive geometrical chirality index, as calculated according to
the definition in Ref.~\onlinecite{Potts}. We define it as a right
handed gammadion (RHG). The gammadion opening in the screen has
fourfold rotational symmetry around an axis perpendicular to the
screen and a plane of symmetry that is parallel to the screen. We
investigated a free-standing flat gold film of thickness $d =$ 140
nm with a single gammadion shaped slit cut into it.  The width of
the gammadion slit was chosen to be $w=$ 0.6 $\mu$m, while the
total length of the gammadion, measured from the end of one arm to
the end of the opposing arm, was $L=$ 4.1 $\mu$m
(Fig.~\ref{Fig1}(a)). For our analysis we used realistic values of
the complex dielectric coefficient of gold at different
wavelengths from Ref.~\onlinecite{Palik}. To evaluate the
propagation of light through the metal screen with its chiral
opening we numerically modelled the electromagnetic fields at a
distance of 160 nm from the screen on the opposite side to a
normally incident plane monochromatic electromagnetic wave. In our
analysis we used the FEMLAB software package that implements a
true 3D finite element method for solving Maxwells equations
\cite{Jin}.  We studied the transmission of the chiral hole for
light at wavelengths $\lambda$ from 1.3 $\mu$m to 5.6 $\mu$m, thus
covering wavelengths both shorter and longer than the
characteristic size of the gammadion opening $L=$ 4.1 $\mu$m. All
results are presented in terms of the Stokes parameters of the
incident $S^{in}_i$ and transmitted $S^{tr}_i$ fields that are
defined in a standard fashion \cite{Svirko,Jackson} in a Cartesian
coordinate frame. In this case $S_0$ is proportional to the light
intensity, while the combinations $(S_0 \pm S_3)/2$ represent the
intensities of the left- and right-handed polarization components
of the field.

We found that the light transmitted through the structure accrues
a polarization component opposite to the polarization state of the
incident wave, giving rise to a circular polarization conversion
effect. Thus the transmitted field becomes elliptically polarized,
with its polarization state being a function of coordinate
$\textbf{r}$ in the the $xy$ plane. Figure~\ref{Fig1} illustrates
a dramatic polarization conversion at $\lambda$ = 4.1 $\mu$m. The
observed polarization conversion effect with circularly polarized
light is governed by chirality. It is different by its
manifestation and nature from the well documented polarization
conversion effect for linearly polarized light that is driven by
anisotropy of the structure and may be seen in a metal grating
consisting of a series of high and narrow ridges that are oriented
at 45 degrees to the polarization angle \cite{Hooper}. The
dependence of the effect reported here on the handedness of a
circularly polarized wave is obvious when comparing the intensity
maps of transmitted light with converted polarization for
illumination with a RCP plane wave (Fig.~\ref{Fig1}(b)) and a LCP
plane wave (Fig.~\ref{Fig1}(c)). These field maps are dramatically
different. The effect of enantiomeric sensitivity is also
illustrated by Fig.~\ref{Fig1}(d) and Fig.~\ref{Fig1}(e), which
show that total transmission is sensitive to the mutual handedness
of wave and gammadion hole.
\begin{figure*}
    \includegraphics[width=180mm]{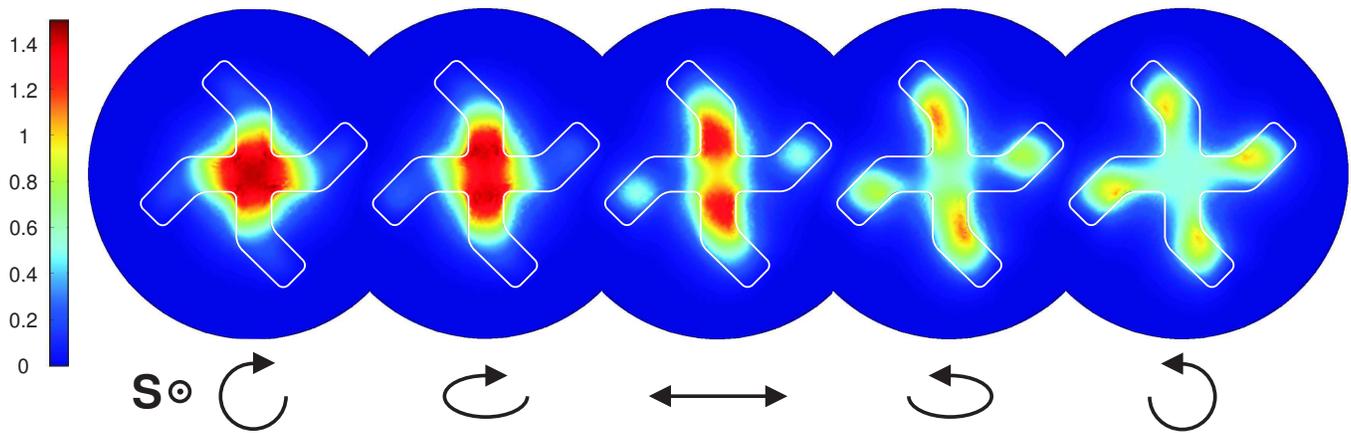}\\
  \onecolumngrid
  \caption{\label{Fig3}Dependence of the filed structure behind the gammadion hole on the polarization state of incident
  light indicated by arrows.}
\end{figure*}

By integrating the intensities of the two circularly polarized
components over the whole gammadion opening we calculated the
overall power transmission  $T=\int S^{tr}_0 (\textbf{r}) d\sigma
/ \int S^{in}_0 (\textbf{r}) d\sigma$ and the overall polarization
conversion efficiency $K_{\pm}=\int (S^{tr}_0 (\textbf{r}) \pm
S^{tr}_3 (\textbf{r})) d\sigma / \int 2 S^{in}_0 (\textbf{r})
d\sigma$ of the screen with chiral hole at different wavelengths
(here $d \sigma$ is an area element of the screen surface). The
wavelength dependence of the overall power transmission of the
hole (Fig.~\ref{Fig2}(a)) shows only a small variation with the
wavelength. Within the accuracy of our numerical method we found
no dependence of the overall power transmission on the incident
polarization state. However, chiral effects are very pronounced in
polarization conversion, as can be seen from Fig.~\ref{Fig2}(b).
When the handedness of the chiral hole and the wave are opposite
(solid line on Fig.~\ref{Fig2}(b)), the polarization conversion
efficiency reaches a maximum at a wavelength close to the
characteristic size of the chiral opening, $\lambda \simeq L$.
When handedness of the chiral hole and the wave coincide, the
polarization conversion is nearly a factor of two less efficient
and has a maximum shifted to a shorter wavelength (dashed line on
Fig.~\ref{Fig2}(b)). The chiral sensitivity diminishes for
incident light wavelengths $\lambda$ much shorter than the
characteristic size of the chiral opening $L$. This is
demonstrated by the convergence of solid and dashed lines in
Fig.~\ref{Fig2}(b) at $\lambda<1800$ nm. If these numerical
experiments are repeated with a left-handed opening (a mirror
image of the structure in Fig.~\ref{Fig1}), the polarization
conversion effect becomes more efficient for right handed
circularly polarized light. From here one can make the intriguing
observation that since the perceived sense of twist of the chiral
hole depends on what side the screen is observed from, the
polarization conversion efficiency depends on whether the light
wave enters the screen from one side or the other (see the inset
in Fig.~\ref{Fig2}).

We observed another interesting and potentially useful phenomenon
which may be seen on Fig.~\ref{Fig3}. The light field behind the
screen is strongly concentrated at the center of the gammadion
when handedness of the chiral hole and the polarization state of
the wave are the same. Here light energy is concentrated in a spot
with a diameter much smaller than both the wavelength and the
characteristic size of the gammadion. In contrast, for opposite
incident polarization no concentration effect is seen and the
transmitted light energy is spread across the whole opening.

In summary we have found polarization conversion and nano-focusing
of circularly polarized light transmitted through a chiral hole in
a metallic screen. Both depend on the mutual sense of twist of the
hole and the handedness of the incident wave. They are pronounced
for wavelengths close to the overall size of the opening and
disappears for shorter wavelengths.

The authors  acknowledge the support of the Engineering and
Physical Sciences Research Council (UK) and fruitful discussions
with A. Zayats.


\end{document}